\documentclass[12pt]{article}
\usepackage{epsfig,enumerate,verbatim}
\usepackage{graphicx}
\begin{document}

\begin{center}
{\large \bf Nonlinear Quantum Cosmology of de Sitter Space}
\end{center}
\vspace{0.1in}

\begin{center}

{Rajesh R. Parwani\footnote{Email: parwani@nus.edu.sg} and Siti Nursaba Tarih\footnote{Email: a0076582@nus.edu.sg}}

\vspace{0.3in}

{Department of Physics,\\}
{National University of Singapore,\\}
{Kent Ridge,\\}
{ Singapore.}

\vspace{0.3in}

\end{center}
\vspace{0.1in}
\begin{abstract}
We perform a minisuperspace analysis of an information-theoretic nonlinear Wheeler-deWitt (WDW) equation for de Sitter universes.  The nonlinear WDW equation, which is in the form of a difference-differential equation, is transformed into a pure difference equation for the probability density by using the current conservation constraint.  In the present study we observe some new features not seen in our previous approximate investigation, such as a nonzero minimum and maximum allowable size to the quantum universe: An examination of the effective classical dynamics supports the interpretation of a bouncing universe. The  studied model suggests implications for the early universe,  and plausibly also for the future of an ongoing accelerating phase of the universe.   
\end{abstract}

\vspace{0.5in}

\section{Introduction}

While quantum physics is expected to play an important role near the Big Bang when the universe was small, there are a number of differing opinions on how to quantise the classical universe, see reviews in \cite{QG, LQC}, leading to different approaches to quantum cosmology. One issue of interest in quantum cosmology is whether classical singularities are resolved \cite{Sing}. In our previous study \cite{NP1} we found the answer to be affirmative in the information-theoretic approach, as is the case also in many other approaches \cite{QG, LQC, Sing}.

The philosophy of the information-theoretic  approach, also known as the ``maximum uncertainty (entropy) method", is that one should minimise any bias when choosing probability distributions, while still satisfying relevant constraints \cite{Jaynes}. While this approach originated in statistical mechanics, it has wider applicability such as motivating the structure of the usual Schrodinger equation \cite{Sch} and its potential nonlinear modifications to model new short distance effects\cite{RP1, Tabia, RP2}.  

In our previous study \cite{NP1}, which we briefly review in the next section, we treated the nonlinearity of the modified WDW equation (\ref{nl1}) perturbatively and studied a truncated linearised equation with an effective potential. In this paper we present a full investigation for the case of a de Sitter universe. The de Sitter universe is a reasonable model for early stages of inflation and furthermore, if the current accelerating phase of our universe continues, then at late times it can again be well approximated as a de Sitter universe. 

Treating the universe as an isolated system, one may apply quantum mechanics to the whole de Sitter universe regardless of its size; in other words, study the wavefunction of the universe. Our main intent in this paper is to see how the classical de Sitter dynamics is modified by information-theoretically motivated corrections to the WDW equation, extending the previous \cite{NP1} approximate investigation. 


Consider the Einstein-Hilbert action for a  FRW universe with the cosmological constant $\Lambda =3/a_{0}^2$ modeling inflationary sources in the early universe, 
\begin{equation}
S=\int dt L = {1 \over 2} \int dt N \left[ {-\dot{a}^2 a\over N^2 } + a(k- {a^2 \over a_{0}^2} ) \right] \label{frw1} \, .
\end{equation}
where  $0 \le a(t) < \infty$ is the dimensionless\footnote{As in \cite{NP1}, the physical scale factor $a_{phys}=\beta a$, where $\beta \propto l_p$, the Planck length.} scale factor,  $N$ the lapse function, $k=0,\pm 1$, and we have taken $\hbar=c=1$.  Varying with respect to $N$ and choosing the $N=1$ gauge gives us the Friedmann equation  
 \begin{equation}
 \dot{a}^2 + (k-a^2) =0 \, . \label{fried}
\end{equation}
We have set  $a_{0} =1$ as the results for other values can be obtained by scaling. 

The flat, $k=0$, geometry has the expanding classical solution
\begin{equation}
a=\exp(t) \label{k0}
\end{equation}
which implies an arbitrarily small universe, $a \to 0$, at early times $ t \to -\infty$. Expected quantum effects are studied in standard minisuperspace quantisation by promoting the canonical momentum $\Pi={\partial L \over \partial \dot{a}} =-\dot{a}a$ to an operator, $\Pi \to \hat{\Pi} = -i {\partial \over \partial a}$, leading to the WDW equation  \cite{Atkatz},
\begin{equation}
\left[ -{\partial^2 \over \partial a^2}  -a^4 \right] \psi(a) =0 \, . \label{wdw1}
\end{equation}

The solution of (\ref{wdw1}) representing an expanding universe is given by the Hankel function
\begin{eqnarray}
\psi_0(a) &\propto&  \sqrt{a} H_{1/6}^{(2)} (a^3/3) \label{out} \\
&\sim& \sqrt{{6 \over \pi a^2}} \exp{\left[ -i(a^3 -\pi)/3\right]} \,\,\,  \mbox{as} \; \; a \to \infty \, , \label{asymp}
\end{eqnarray}
as one can verify by noting that the negative value for the momentum $\Pi=-a \dot{a}$ of the asymptotic form corresponds to   $\dot {a} >0$.  
Notice that the , solution (\ref{out}) of the linear WDW gives a nonzero probability for a universe of size $a=0$, and hence for a Big Bang. In a previous perturbative study \cite{NP1} we showed that a modified WDW dynamics screened the $a=0$ region, leading to the quantum creation of a universe through tunneling \cite{Atkatz}.

We postulated in Ref.\cite{NP1} that if there is some new physics at quantum gravity scales, it may be modeled, within the information theory framework, by a modified WDW equation \cite{RP1} which can be derived by extremising the usual WDW Lagrangian while also maximising the relevant uncertainty measure. We state and explain first the equation before motivating it further below: 
\begin{equation}
\left[ -{\partial^2 \over \partial a^2} -a^4 + F(p) \right] \psi(a) =0 \label{nl1}
\end{equation}
where
\begin{eqnarray} 
  F(p) &\equiv& Q_{NL} - Q \, , \label{F} 
\end{eqnarray}
with 
\begin{equation}
Q_{NL}= {  1  \over 2 \zeta^2 \eta^2}  \left[ \ln {p \over (1-\eta) p + \eta p_{+} } +  {\eta p_{+} \over (1-\eta) p + \eta p_{+}} - {\eta p_{-} \over (1-\eta) p_{-} + \eta p} \right]  \,  \label{Q2}
\end{equation}
and 
\begin{eqnarray}
Q &=&  -  {1 \over \sqrt{p}} {\partial^2 \sqrt{p} \over \partial a^2} \, \; . \label{pot1} 
\end{eqnarray}
Here $p(a) = \psi^{\star}(a) \psi(a)$ is the probability density and $p_{\pm}(a)  \equiv  p(a \pm \zeta)$, where the dimensionless parameter  $\zeta>0$ is the  nonlinearity scale\footnote{We have changed notation from that in Ref.\cite{NP1}.}. The linear theory is recovered as $\zeta \to 0$  while the other parameter  $0 < \eta < 1$ labels a family of nonlinearisations. Notice that the modified WDW equation is still invariant under a scaling of the wavefunction, $\psi \to \lambda \psi$, and so the solutions of the equation do not depend on the normalisation of the probability density. 

The appearance of $p_{\pm}$ in (\ref{nl1}) implies a nonlocality of the modified WDW equation which takes the form of a difference-differential equation.
While the equation itself looks complicated, it follows from a relatively simple Lagrangian involving the Kullback-Leibler (KL) information measure \cite{RP1,NP1}: The nonlinear piece $F$ in Eq.(\ref{nl1}) arises from maximising, in the spirit of the maximum uncertainty method, the following term in the action 
\begin{equation}
I_{KL}(p,p_{+}) \propto - \int p(a) \ln {p(a) \over p(a+\zeta)} \ da \label{kl} \, .
\end{equation}

The Kullback-Leibler measure is clearly a relative uncertainty measure which generalises the usual Gibbs-Shannon entropy of statistical mechanics, and which reduces, as $\zeta \to 0$,  to the ``Fisher" information measure used in Ref.\cite{Sch} for motivating the usual Schrodinger equation within the information-theoretic framework. Indeed, as discussed in \cite{RP1}, the ($\eta$-regularised) KL measure is probably the simplest nonsingular measure which interpolates between the Shannon and Fisher measures and which keeps some desirable properties such as the scale invariance mentioned above. 

In physical terms, the nonlinearity scale $\zeta$ may be interpreted as a kinematic implementation of the resolution at which the coordinates become distinguishable \cite{RP1}. Clearly for  $\zeta \neq 0$ the  dynamics given by Eq.(\ref{nl1}) will  be modified from the usual case. 

The nonlinearity in (\ref{nl1}) was originally explored in quantum mechanical systems \cite{RP1}, and its various perturbative and non-perturbative properties studied in Refs.\cite{RP1,Tabia,RP2,work}. However as experimental constraints simply place limits on the size of the nonlinearity for simple quantum mechanical systems \cite{RP1}, we then applied the modified quantum equation to cosmology in Ref.\cite{NP1} and found encouraging results for singularity resolution.  In this paper we continue our study of the consequences of the information-theoretically motivated nonlinear WDW equation (\ref{nl1}).

The outline of this paper is as follows. In the next section we review the perturbative results of Ref.\cite{NP1} on how the nonlinearly corrected quantum dynamics can avoid the $a=0$ possibility that is present in Eq.(\ref{out}). Then in Section(3) we show how to transform (\ref{nl1}) into a purely difference equation for $p$ which is solved numerically in Section(4) and studied analytically in Section(5). In Section(6) we discuss the effective classical
dynamics suggested by the nonlinear WDW equation and elucidate the physical meaning of the nodes of the wavefunction. In Section(7) we discuss the $\Lambda =0$ case and conclude in Section(8).

\section{Review of perturbative treatment}
If the nonlinearity  $F$ is weak, it may be expanded perturbatively for $\zeta \ll 1$, giving to lowest order
\begin{equation}
F(p) = \zeta(3-4\eta) f(p) + O(\zeta^2) \, , \label{expand}
\end{equation} 
where
\begin{equation}
f(p)={p' \over 12 p^3} (2 p'^2 -3p''p) \, .  \label{f1est}
\end{equation}
In this approximation Eq. (\ref{nl1}) is  
\begin{equation}
\left[ -{\partial^2 \over \partial a^2} - a^4 + \zeta(3-4\eta) f(a) \right] \psi(a) =0 \label{trunc}
\end{equation}
which we may solve  by iterating about the unperturbed solution (\ref{out}): At lowest order one calculates $p_0=\psi_{0}^{*} \psi_{0}$ and then $f_0(a)$, giving a linear Schrodinger equation with an effective potential 
\begin{equation}
V_{eff} = -a^4 + \zeta(3-4\eta) f_0(a)  \, . \label{iter}
\end{equation}

The perturbative approximation (\ref{expand}) around the linear solution requires not just $\zeta \ll 1$ but also that $f_0(a)$ be slowly varying, which is indeed the case since for $a\to \infty$ we have $f_0 \sim 1/a^3$.  In particular, there are no singularities from nodes of the wavefunction in the expansion of $F(p)$  around $p_0$, unlike the case of quantum mechanical systems studied in \cite{Tabia}.

As shown earlier in \cite{NP1}, for $\eta <3/4$ the nonlinearity $f_0(a)$ forms an effective potential barrier, a finite size universe coming into being through quantum tunneling \cite{Atkatz}. In other words, in the modified classical dynamics a backward evolving classical universe will experience a bounce instead of shrinking to zero size.

In the sections below we venture beyond the perturbative analysis of Ref.\cite{NP1} to uncover some new features of Eq.(\ref{nl1}).

\section{Exact difference equation for $p$}
Writing the exact wavefunction in terms of its amplitude  and phase, $\psi = \sqrt{p} e^{iS}$, with $p$ and $S$ real, the imaginary part of the WDW equation (\ref{nl1}) is then the continuity equation
\begin{equation}
{\partial \over \partial a} \left( p {\partial S \over \partial a} \right) =0 \, ,
\end{equation}
which can be solved to give  
\begin{equation}
p {\partial S \over \partial a} = \sigma  \label{ds} \, .
\end{equation}
The constant current $\sigma$ is fixed by requiring our nonperturbative solution approach the asymptotic form\footnote{However as we shall see, for even larger $a$, beyond the matching point, the nonlinear WDW solution will eventually deviate from the form (\ref{asymp}) indicating an eventual departure from the current classical evolution.} of the solution for the linear theory (\ref{asymp}) near some large $a$, which corresponds, for example, to ``now". This gives
\begin{equation}
\sigma =  {-6 \over \pi} \, .
\end{equation}
Next, we use (\ref{ds}) in the real part of the nonlinear WDW equation (\ref{nl1}) to eliminate the derivatives of $S$, giving a purely difference equation for the probability density:
\begin{eqnarray}
\left({\sigma \over p}\right)^2 &=& a^4 - Q_{NL} \label{diff} \\
&=& a^4 - {  1  \over 2 \zeta^2 \eta^2}  \left[ \ln {p \over (1-\eta) p + \eta p_{+} } +  {\eta p_{+} \over (1-\eta) p + \eta p_{+}} - {\eta p_{-} \over (1-\eta) p_{-} + \eta p} \right] \nonumber \, .
\end{eqnarray}

In the derivation we factored a common $\sqrt{p}$ from both sides of (\ref{diff}); as discussed at the end of Sect.(5.1), this does not affect the results even as $p \to 0$. The difference equation (\ref{diff}) relates the adjacent values of probability density $p_{-}, p $ and $p_{+}$ which are separated by the step size $\zeta$, the nonlinearity scale


It is important to note that the variable $a$  is still continuous, as is $p(a)$. It is just that equation (\ref{diff}) places non-local constraints on the $p(a)$ values. Thus although the solutions to be discussed below are on a lattice of step size $\zeta$ as determined by (\ref{diff}), it is to be understood that the region between the discrete set of points is continuously connected.

The equation (\ref{diff}) for the {\it probability density} does not depend on the sign of $\sigma$ and hence describes both possibilities, either an expanding or contracting universe; the specific {\it wavefunction} describing either possibility does depend on $\sigma$. 

As $\zeta \to 0$, $-Q_{NL}$ becomes $-Q >0$, Eq.(\ref{pot1}), which is positive definite for the linear WDW equation solution (\ref{out}) and hence the right-hand side of (\ref{diff}) becomes positive definite, just as the left-hand side already is. However,  for $\zeta \neq 0$ the difference equation (\ref{diff}) restricts the range of $a$ as we are required, by definition, to preserve the positivity of the probability density. When starting with $p(a)>0$ at some initial point and using the difference equation to move forward or backward, it is possible that one reaches a point at which $p=0$, beyond which $p<0$ or becomes complex, as the equation (\ref{diff}) by itself does not guarantee a real or positive $p_{\pm}$: the kinematic constraint $p>0$ must be self-consistently imposed. 

In Sects.(4,5) we interpret the occurrence of $p=0$ as delimiting the range of allowed $a$ values and hence on the size of the universe: The consistency of such an interpretation will be  seen by examining the effective classical dynamics in Sect.(6).  

In summary, the kinematic physical constraint 
$p>0$, when imposed on the dynamical difference equation (\ref{diff}), constrains the size of the universe. (Such a constraint does not occur for usual quantum mechanical systems, see later).

\section{Numerical Results}

The difference equation (\ref{diff}) is easily solved by specifying two initial values which we fix by requiring the large universe to be close to the asymptotic form (\ref{asymp}) given by the usual linear WDW equation near $a=5$. Since Eq.(\ref{diff}) easily gives $p_{-}$ explicitly in terms of $p$ and $p_{+}$, hence if the later two variables are initially fixed then a direct backward evolution  of the equation gives the values of $p$ for smaller $a$. The numerical accuracy was set at $16$ figures which allows the features of $p$ discussed below to be unambiguously distinguished. 

However in Eq.(\ref{diff}) $p_{+}$ cannot be written explicitly in terms of the other two values, so values of the probability density  forward from the starting point were obtained by solving the implicit equation using Newton's method. We checked the accuracy of Newton's method by making several consistency comparisons; for example, using the end points of the forward steps as starting points for the direct backward evolution and comparing the two curves. Identical results were obtained using the ``Solve" function in MATLAB. 

 We summarise below the key results. Unless otherwise stated,  $\eta=0.5$. For the nonlinearity scale, we explored the range $10^{-3} <\zeta < 1 $ though not with the same degree of detail for every feature. 

\begin{enumerate}

\item For low nonlinearity, $\zeta \to 0^{+}$, starting from $a=5$ and moving backwards towards $a=0$, $p$ remains positive and close to the $p$ for the linear theory. These results, including the effective potential $V_{eff} \equiv  -a^4 + F(p)$ which develops a barrier, Fig.(1), agree with those obtained using the lowest order perturbation theory in Ref.\cite{NP1}; the barrier being smaller for smaller $\zeta$. (For the evaluation of the effective potential using the discrete $p$ data we used the central difference approximation for the second derivative in $Q$ (\ref{pot1}), which is sufficient for the low $\zeta$ values encountered.)

However, moving forward from the starting point, the probability density will eventually become zero at some\footnote{In the numerical work we locate the first point where $p$ becomes negative or complex. Due to the lattice nature of (\ref{diff}), the actual point where the probability density first vanishes will be between two lattice points.} finite value which we label as $a_{max}$. The consistency of interpreting $a_{max}$ (and $a_{min}$ below) as their labels suggest will be seen in Sect.(6). The value $a_{max}$ depends on $\eta,\zeta$, for example $a_{max} = 14.4$ for $\eta=1/2$ and $\zeta=0.005$ for the initial conditions used; Figures (2,3) show the curves. The general trend is that $a_{max}$ decreases with increasing $\zeta$ for $\zeta < \zeta_c$ (as defined below). 

Thus, unlike the perturbative study in Ref.\cite{NP1}, the non-perturbative results show a maximum allowable size to the quantum universe. For low values of $\zeta$, $a_{max}$ fits a power law in $\zeta$, Fig.(4). 

\item As the nonlinearity $\zeta$ increases beyond a critical value $\zeta_c$, $p$ becomes zero at some $a_{min}(\eta,\zeta)$ during the backward evolution, in addition to vanishing at a forward point $a_{max}(\eta,\zeta)$ during the forward evolution. For example, $\zeta_c=0.049$ for $\eta=0.5$ and the initial conditions used, Fig.(5). In the range $\zeta \gg \zeta_c$ the trend is that increasing $\zeta$  leads to increasing $a_{max}$ and decreasing $a_{min}$.
 
The existence of $a_{min}$ means\footnote{In some cases, a continued evolution beyond $a_{min}$ or $a_{max}$, where $p$ is negative or complex, leads to new regions where $p$ becomes positive again. We assume that the wavefunction must be continuous and so display only the positive component of $p$ around the initial point.} that the quantum universe has a nonzero minimal allowable size: This is the second new result not seen in the perturbative treatment of Ref.\cite{NP1}. Such a possibility, of a minimum size to space determined by the quantum nonlinearity $F$, was earlier noted in Ref.\cite{RP2}.  


\item As $\zeta$ increases from zero, oscillations become apparent in the probability density at larger values of $a$, in contrast to the probability density for the linear theory (\ref{out}) which is monotonically decreasing. These oscillations are discussed in more detail in the next Section.

\item We checked that the new features observed in our numerical study of (\ref{diff}) are robust to changes in initial values for  $p$ or the starting point $a$ used. For example, we solved the equation for starting points near $a=10$ and used starting values for two of the adjacent $p$'s in (\ref{diff}) to be the same small value; the results again show the existence of $a_{min}, a_{max}$ and the oscillations though the specific numerical values differ.

\item All the above results are qualitatively the same for a lower  value $\eta=0.2$ that we checked. However for larger $\eta=0.9$, though there is still an $a_{max}$ and a $\zeta_c$, there is no potential barrier at small $\zeta$ in agreement with Ref.\cite{NP1} which showed the barrier to be present at low $\zeta$ only for $\eta < 0.75$.

\end{enumerate}

\section{Analysis}

The features observed in the numerical study may be understood through various analytical approximations which we discuss in this section.

\subsection{Existence of $a_{min}$ and $a_{max}$}

As an illustration, a relation for $a_{max}$ can be estimated for $\zeta$ small as follows: Set $p_{+}=0$ at $a=a_{max}$ in (\ref{diff}) and as $\zeta \to 0$ estimate $\psi=\psi(a_{max}-\zeta) \approx -\zeta \psi^{'} \equiv -\gamma \zeta$ so that $p \approx (\gamma \zeta)^2$. Similarly $p_{-} \approx (2 \gamma \zeta)^2$ where $\gamma$ is the slope\footnote{We are assuming here that it is the wavefunction which is smooth near the node. If instead the probability density is smooth then an expression different from but similar to Eq.(\ref{amax}) can be derived.} of the wavefunction at $a$. Then 
\begin{equation}
a_{max} \approx {1 \over \sqrt{\zeta}} \left( {\sigma^2 \over (\gamma^2 \zeta)^2 } - {1 \over 2 \eta^2} ( \ln(1-\eta) + 
{4 \eta \over 4-3\eta} ) \right)^{1/4} \label{amax}
\end{equation}
Note that the slope $\gamma$ is in general a function of $\zeta$ and $a_{max}$ but if that dependence does not overcome the explicit $\zeta$ factors in (\ref{amax}) then we see that $a_{max} \to \infty$ as $\zeta \to 0$. 
The numerical results for $a_{max}$ are shown in Fig.(4). For $\zeta$ small we find  $a_{max} \propto {1 \over \sqrt{\zeta}}$.

While setting $p_{+}$ or $p_{-}$ to zero in Eq.(\ref{diff}) is unproblematic, it appears that $p \to 0$ leads to a divergence through the $\ln(p)$ term. However re-arranging $(\ref{diff})$ shows that $p \to 0$ leads to the following consistency relation
\begin{equation}
-2\eta^2 \zeta^2 \sigma^2 = \lim_{p \to 0} p^2 \left[ {\eta p_{+} \over (1-\eta) p + \eta p_{+}} - {\eta p_{-} \over (1-\eta) p_{-} + \eta p}\right]
\end{equation}
which implies that the expression in square-brackets must develop a $-1/p^2$ divergence through either $p_{+}$ or $p_{-}$ becoming negative. That is, as $p \to 0$, one of the adjacent points enters the unphysical region beyond either $a_{max}$ or $a_{min}$. As discussed earlier, this may be interpreted as implying that the quantum universe within this nonlinear WDW framework is bounded.\footnote{The situation is different for the analogous difference equation for bounded potentials in quantum systems \cite{RP1,Tabia,RP2,work}: In those cases such a constraint does not arise.}.

\subsection{An Exact Solution}

An exact analytic solution of (\ref{diff}), which illustrates the occurrence of both an $a_{min} >0$ and $a_{max}$, is obtained by taking three lattice points with $p_{-} = p_{+}=0$ and $p \equiv p(a_m) >0$ at the mid-point $a_m$. Then (\ref{diff})  gives $p$ in terms of $a_m, \zeta, \eta$.
\begin{equation}
{\sigma^2 \over p^2} = a_{m}^{4} - { 1 \over 2 \eta^2 \zeta^2} \ln {1 \over 1-\eta}  \, .\label{ex1}
\end{equation} 
Since the right-hand side of (\ref{ex1}) must be non-negative, this sets one constraint. Also since $a_m - \zeta = a_{min}>0$ this gives the second constraint $a_m = \zeta + a_{min}$. Thus (\ref{ex1}) implies 
\begin{equation}
\zeta^2 \ (\zeta + a_{min})^4  > { 1 \over 2 \eta^2} \ln {1 \over 1-\eta} \, .
\end{equation}
One can have large universes, up to size $a_{min} + 2 \zeta$,  by taking $\eta \to 0^{+} \, \mbox{or} \, 1^{-}$.

\subsection{Oscillations in the Probability Density}
Return to the nonlinear WDW equation (\ref{nl1}) and now write the the wavefunction as $\exp{(\theta/2 + iS)}$ for real $\theta,S$ and treat the nonlinearity $F$ as one piece rather than separating $Q_{NL}$ from $Q$. As in Sect.(3), one may eliminate $S$ and obtain the following equation,
\begin{equation}
2 \theta^{''} + (\theta^{'})^2 + 4( a^4 - \sigma^2 e^{-2 \theta} ) + \epsilon =0 \, , \label{TE}
\end{equation}  
where $\epsilon$ represents the terms from the nonlinearity $F$ which we treat in the analysis below as small perturbations to the original linear WDW equation; the prime means $d/da$. Then, if $\theta_0$ is the solution to 
 Eq.(\ref{TE}) for $\epsilon=0$, the full solution may be written $\theta = \theta_0 + \delta$ and this substitution in (\ref{TE}) gives, to lowest order in $\epsilon$ and $\delta$, the equation for the fluctuations  
\begin{equation}
 \delta^{''} +  \theta_{0}^{'} \delta^{'} + 4 \sigma^2 \delta e^{-2 \theta_{0}} = 0 \, . \label{fluc}
\end{equation}
Since the probability density $p_0$ corresponding to the solution of the linear WDW (\ref{wdw1}) is monotonically decreasing, so $\theta_{0}^{'} <0$ and hence the fluctuation equation (\ref{fluc}) describes oscillations which increase with $a$ (anti-damping). As $p_0 \propto 1/ a^2$ for large $a$, therefore $\theta^{'}_{0} \propto -1/a$ and the anti-damping eventually vanishes. Furthermore, the last term of (\ref{fluc}) implies that the wavelength of oscillations decrease with $a$. 

This analysis explains why in our numerical solutions the oscillations were not visible at low $a$: their initial amplitude was small and the wavelength large. The oscillations became manifest only as their amplitude increased and wavelength decreased at larger $a$; though the amplitude eventually stabilises, the wavelength keeps decreasing\footnote{Notice that the leading order analysis does not refer to the specific form of $F$. Any small perturbation of the original linear WDW equation is expected to give rise to qualitatively similar oscillations in the probability density.}, see Fig.(6).  

Including sub-leading terms in (\ref{fluc}) would give the oscillations a dependence on the parameters $\eta, \zeta$ as we have also observed numerically. As $\delta \to 0$ when $\zeta \to 0$, the amplitude of the oscillations must also be of order $\zeta$ or smaller. The ratio $p/p_0 = \exp(\delta) \sim 1 + \delta$ then implies small oscillations about the value $1$ as illustrated in the actual numerical results of Fig.(5).

We have also studied the truncated differential equation (\ref{trunc}) obtained by expanding $F$  to lowest order in $\zeta$. Direct solution of that nonlinear differential equation shows oscillations for $p$, as does the first order iterative version  described by the linear Schrodinger equation with effective potential (\ref{iter}). However neither of these approximations to Eq.(\ref{diff})   display the limiting values $a_{max}, a_{min}$  seen in the nonperturbative difference equation (\ref{diff}). 

The oscillations we have observed here do not seem to be related to the oscillations of the non-perturbative mode of a similar nonlinear Schrodinger equation studied in Ref.\cite{RP1}: the latter oscillations   were of constant wavelength.

\section{Effective Classical Dynamics}

As the WDW equation is time-independent, it  cannot describe the time-evolution of a universe. Furthermore the probability density refers to the likelihood of observing a universe of a particular size in an ensemble. Therefore, in order to gain some 
insight into the dynamics of a single universe, in particular near $a_{max}$ and $a_{min}$, we return to the classical domain. Near the nodes, we see from (\ref{ds}) that $dS/da$ is large and as argued in Refs.\cite{Halliwell, Atkatz} it may be identified with the classical momentum, as in the discussion after Eq.(\ref{asymp}).  

In that approximation the effective classical dynamics corresponding to (\ref{nl1}) is  described by the modified Friedmann equation
\begin{equation}
a^2 \dot{a}^2  + V_{eff} = 0  \label{modF}
\end{equation}
where $V_{eff} = -a^4 + F(p)$  is the effective potential and $p$ is the solution of the nonlinear WDW equation. Note that the modified Friedmann equation is again a difference-differential equation. In Ref.\cite{NP1} we treated (\ref{modF}) perturbatively by expanding $F$ to lowest order in $\zeta$ but in the discussion below we keep the full form of $F$.

If the nonlinearity $F$ is weak and non-singular then it cannot overcome the classical piece $-a^4$, especially at large $a$. However near $a=a_{*} >0$, where the wavefunction vanishes, $p \sim (a-a_{*})^2$ and so from Sect.(5.1) we see that $Q_{NL} \sim -1/p^2$ near $a_{*}$. Similarly, from Eq.(\ref{pot1}), $Q$ is also likely to be large and diverging, thus making $F(p) = Q_{NL} -Q$ large and possibly positive. (The enhancement of the nonlinearity near nodes of $p$ was earlier noted in Ref.\cite{Tabia}).

A numerical examination of $V_{eff}$ shows that between $a_{min}$ and $a_{max}$ it is real and negative. In some cases $V_{eff}$ remains real but becomes positive as one approaches either $a_{min}$ or
$a_{max}$ thus forming a potential barrier there which implies, through (\ref{modF}), a bounce: For example, approaching $a_{max}$ from the left, and assuming a Taylor expansion of Eq.(\ref{modF}) near that point, one obtains (dropping positive constants)
\begin{equation}
\dot{a} \approx \sqrt{a_{max} -a}  \, ,
\end{equation} 
showing that $\dot{a} =0$ and $\ddot{a} <0$ as $a \to a_{max}$.  

However for quantum solutions which have both an $a_{min}$ and $a_{max}$, we found that if there is a potential barrier in the effective classical dynamics at one end, say near $a_{min}$, then the effective potential at the other end, near $a_{max}$ is complex. The complex $V_{eff}$ implies that $p$ has reached an unphysical value (negative or complex). Since $Q_{NL}$ is defined only at a discrete set of points, it is possible that $F$ and hence $V_{eff}$ is not analytic near $a=a_{*}$. For example, the  $\ln p \sim \ln(a-a_{*})^2$ term in $Q_{NL}$ might be dominant leading to a transition from a negative to a complex, and hence unphysical $V_{eff}$. In such situations the approximation (\ref{modF}) has broken down near that point. 

 In Ref.\cite{brake}, in the context of a different quantum cosmological model, the authors found that locations of singularities in the classical dynamics were where the wavefunction vanished in the quantum dynamics. In some sense we have found a complementary situation, whereby nodes in the probability density of a nonlinear quantisation implies, in some cases, an effective classical dynamics with  singularities near the nodes.

However it is possible to observe a cyclic universe in the effective classical dynamics if we take $\zeta < \zeta_c$ and $\eta <3/4$, for which there is no $a_{min}$ but still a potential barrier\footnote{We are assuming that the approximation (\ref{modF}) holds also near $a=0$.}   near $a=0$ just as in Sect.(2), and choose $\eta$ such that $V_{eff}$ is real and positive near $a_{max}$. An example is obtained by choosing $\eta=0.7, \zeta=0.02$ which gives an $a_{max}$ near $a=8.1$. Fig.(7) shows $V_{eff}$ near $a_{max}$, showing the positive barrier as one approaches from the left.

In summary, the points $a_{min}$ and $a_{max}$ in the quantum equation imply turning points in the effective classical dynamics (\ref{modF}) in those cases where $V_{eff}$ remain real.  Thus the modified Friedmann equation can support a cyclic universe; for a review on cyclic universes see \cite{cyclic}. Since the difference equation (\ref{diff}) does not depend on the sign of $\sigma$ and hence represents both expanding and contracting solutions, one may argue that the underlying quantum  universe is also cyclic.

Finally, we may also use the effective classical dynamics to interpret the oscillations seen in probability density curves: they imply that  the expansion of the classical FRW-$\Lambda$ universe will deviate from the pure exponential, $a(t) \sim \exp(t(1-\delta))$, with $\delta$ the small oscillatory term discussed in the last section.

\section{$\Lambda=0$ Case}
In a flat universe without a cosmological constant (Minkowski spacetime) the {\it linear} WDW is just the free time-independent Schrodinger equation whose solution is $\psi(a) = c_1 a + c_2$ where $c_i$ are complex constants. The probability density is then $p(a)= Aa^2 -Ba +C$ with $A,C>0$. For $B<0$ the solution is not normalisable but for sufficiently large $B>0$ one gets $p<0$ beyond a certain $a=a_{max}$ and hence an allowed region $0<a<a_{max}$.

When $\Lambda=0$, the $a^4$ piece is absent in the nonlinear difference equation (\ref{diff}) and we set the constant $\sigma =-1$ in (\ref{ds}).  As preliminary numerical trials  indicated that $p$ increased  without bound in the forward evolution if the initial $p$ were nonvanishing, we report the results when we set $p_{+} \approx 0$ at $a=10$ or $a=20$ and choose $p$ to be a small positive value $\sim 0.2$. In Fig.(8) we see also that the nonlinearity makes the $p$ curves less steep compared to that of the linear WDW equation for the same initial conditions at $a_{max}$.

For $\eta=0.2,0.5$ there is no potential barrier formed for small $\zeta$, but just as in the case for $\Lambda \neq 0$ there exists a $\zeta_c$ beyond which the quantum universe has a minimum allowable size, in addition to a maximum size at $a=10$ or $a=20$ that we had set by hand, see Fig.(9).

In passing we note that the exact solution (\ref{ex1}) does not exist for the case $\Lambda=0$ because when  the $a_{m}^4$ piece is absent the equation is inconsistent. However this does not mean that solutions with both an $a_{min} >0$ and $a_{max}$ do not exist when $\Lambda=0$ but rather that one must either consider more than three lattice points, or realise that for a given $a_m$ and $z$ the actual solution might have  $p$ vanishing before the chosen $p_{+}=0$ or after the chosen $p_{-}=0$, as the numerical results do indicate, Fig.(9). In  other words a three lattice point solution may be constructed, for example, by first choosing $p_{-} <0$ and then restricting the solution to the region where the probability density is positive.

\section{Conclusion}
The probability density describing a quantum de Sitter universe in the nonlinear WDW framework of Ref.\cite{NP1} was found to obey a nonlinear difference equation. For an arbitrarily weak nonlinearity, $\zeta$, the universe has a maximum allowable size, $a_{max} \sim {1 \over \sqrt{\zeta}}$, while a potential barrier screens the classical $a=0$ region when the free parameter $\eta <3/4$. 

As $\zeta$ increases beyond $\zeta_c$, which is still very small, there emerges a minimal allowable size $a_{min}$ to the quantum universe for any value of $\eta$. One may think of the quantum nonlinearity as counter-acting \cite{RP2} the dispersion of the wavefunction, localising it to a finite range  $a_{min} < a < a_{max}$.  

In the quantum picture the existence of $a_{min}$ and $a_{max}$ was implied by the vanishing of the probability density: The consistency of interpreting $a_{min}$ and $a_{max}$ as such was seen by examining the effective classical dynamics. We showed that near such nodes of the probability density there correspond bounces in the effective classical evolution for cases where $V_{eff}$ remains real; in some cases  we have a self-consistent interpretation of a cyclic universe. 

While a modified dynamics replacing the Big Bang by a Big Bounce might have been expected, it is remarkable that the evolution of the universe at large scales has also been modified by the nominally weak quantum nonlinearity $F$: Instead of an unending exponential expansion, there can now be a Big Crunch. 
This significant modification of the classical dynamics is due to the enhancement \cite{Tabia} of the nonlinearity  $F$ near the nodes of the probability density. 

The current accelerating phase of our universe is well modeled by a dominant cosmological constant. We interpret our results as suggesting that such an accelerating phase cannot continue indefinitely but will be replaced by a contracting phase. Since our information-theoretic approach does not presume a specific micro-dynamics responsible for the nonlinearity $F$, it might be summarising various possibilities. For example, in Ref.\cite{giddings} a semi-classical analysis showed that in gravitational theories with extra dimensions, such as those that appear in M-theory \cite{daniel}, a vacuum state with non-zero cosmological constant is generically unstable, one possibility being an eventual collapse in a Big Crunch: Our results are compatible with those of \cite{giddings}. In this regard, it is interesting to note that cyclic models of the universe \cite{turok} inspired by M-theory \cite{witten} have been proposed.

Of course to truly see a cyclic time-evolution in the quantum domain of our model we need to go beyond the FRW-$\Lambda$ case and include, for example, a massless scalar field to act as an internal clock \cite{clock}, the FRW-$\phi$ model. The  study of the FRW-$\phi$ model is important since the classical Big Bang has a curvature singularity. 

Based on our perturbative study of the latter model in Ref.\cite{NP1}, and our analysis of the nonlinearity $F$ in Section(5), it is plausible that a non-singular cyclic evolution of the quantum universe might also be seen for the FRW-$\phi$ model within our nonlinear WDW framework \cite{work}. Also, 
a cyclic evolution should also hold for models in which the cosmological ``constant" is slowly varying, in particular for some realistic inflationary potentials \cite{work}. 

Ultimately the free parameters $\eta$ and $\zeta$ need to be fixed by comparing  computations of observables with empirical data. On the other side, the robustness of the theoretical conclusions may be checked by examining information measures which are deformations of the KL measure \cite{RP1}.

Finally we note that the nonlinear difference equation (\ref{diff}) may be viewed as a one-dimensional nonlinear map and could possibly exhibit chaos for certain range of parameters \cite{strogatz}. It might be useful to investigate the physical relevance of such behaviour for the cosmology considered. 


\section*{Acknowledgments}
R.P. thanks Kenneth Ong, Liauw Kee Meng, Saeid Molladavoudi,\\ Panchajanya Banerjee, Tan Hai Siong and Sayan Kar for helpful discussions.

\newpage

\section*{Figure Captions}

\begin{itemize}
\item  Figure 1 : The barrier in the effective potential for $\zeta = 5 \times 10^{-3} < \zeta_c$. (The region below $a=5 \times 10^{-3}$ has been extrapolated since $Q$ cannot be evaluated there using the central difference formula).  

\item  Figure 2: Probability density curve for $\zeta= 5 \times 10^{-3}$. For low $a$ the curve follows very closely the curve for the linear theory but $p$ becomes complex at $a=14.4$.

\item  Figure 3 : The region of Fig.(2) near $a_{max}$.

\item  Figure 4 : Variation of $a_{max}$ with $\zeta$ for $\eta=1/2$.

\item  Figure 5 : The probability density curve for $\zeta =0.1 > \zeta_c =0.049$ showing both an $a_{min}=4.55$ and $a_{max}=5.3$. The discrete data points have been connected for visualisation. Included are the two unphysical points at the ends.

\item  Figure 6 : The oscillations in the $p$ curve shown here in the ratio plot of $p/p_{asymp}$ where $p_{asymp}$ is the asymptotic value for the linear theory, Eq.(\ref{asymp}). 

\item  Figure 7 : $V_{eff}$ near $a_{max}$ for $\eta=0.7, \zeta=0.02$. The last point on the right has imaginary components and is unphysical.  
 
\item Figure 8: Curves for $\Lambda=0$. The upper curve is for the linear WDW equation while the lower is for the nonlinear WDW equation with the same initial conditions near $a= 20$.

\item Figure 9: An example for $\Lambda =0 $ showing both an $a_{min}$ and $a_{max}$. Note the unphysical points at the ends. 
\end{itemize}

\newpage

\section*{Figures}

\begin{figure}[ht]
  \begin{center}
   \epsfig{file=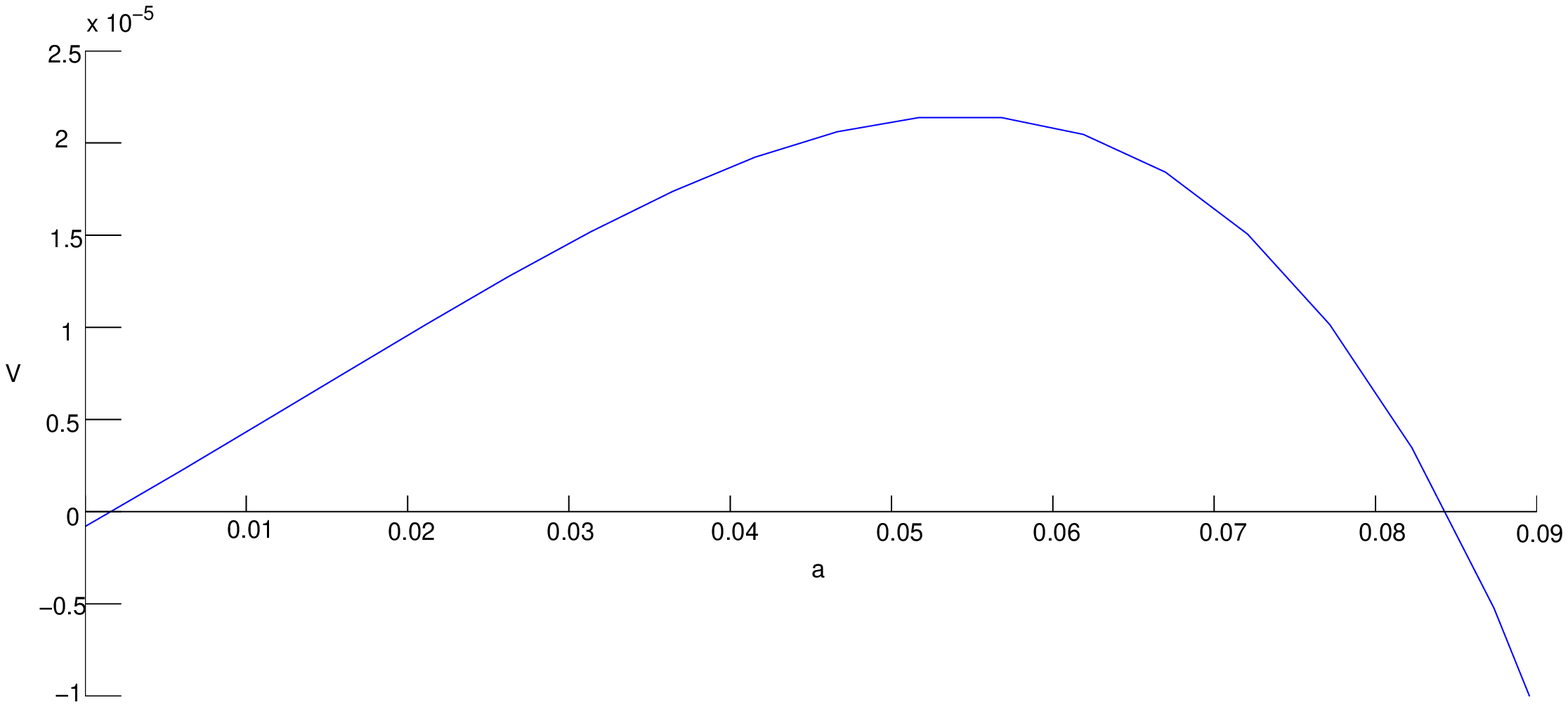, width=12cm}
    \caption{}
  \end{center}
\end{figure}

\begin{figure}
  \begin{center}
   \epsfig{file=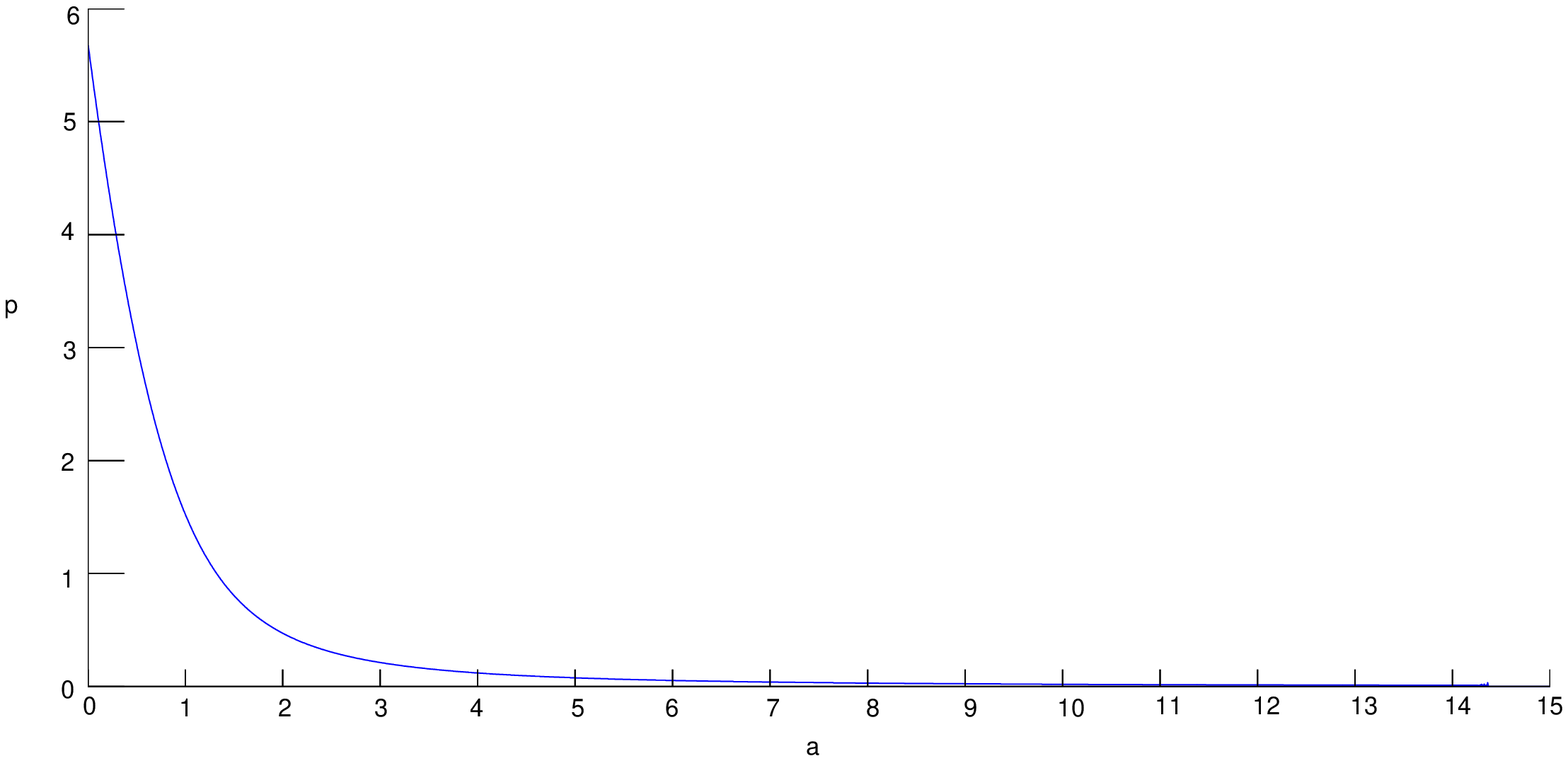, width=12cm}
    \caption{}
  \end{center}
\end{figure}

\begin{figure}
  \begin{center}
   \epsfig{file=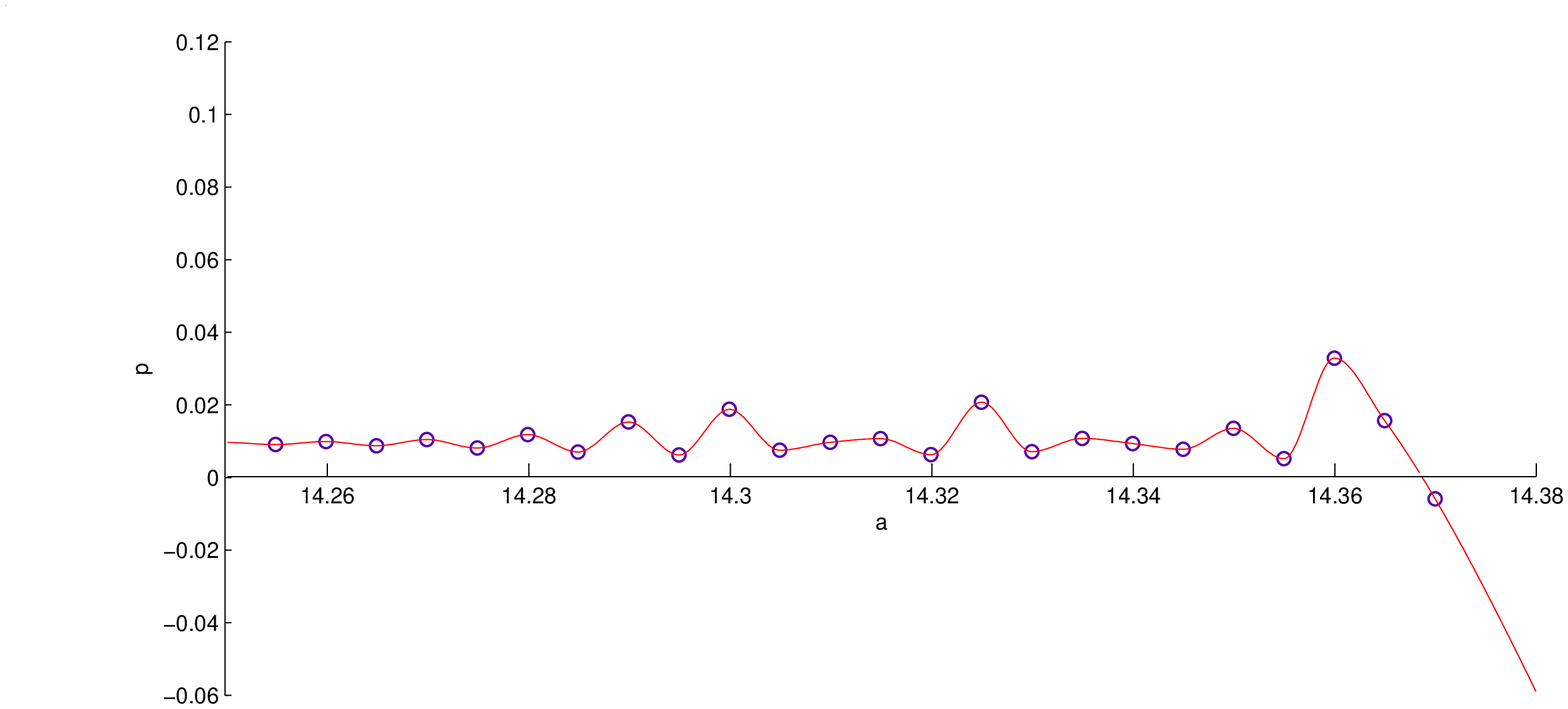, width=12cm}
    \caption{}
  \end{center}
\end{figure}

\begin{figure}
  \begin{center}
   \epsfig{file=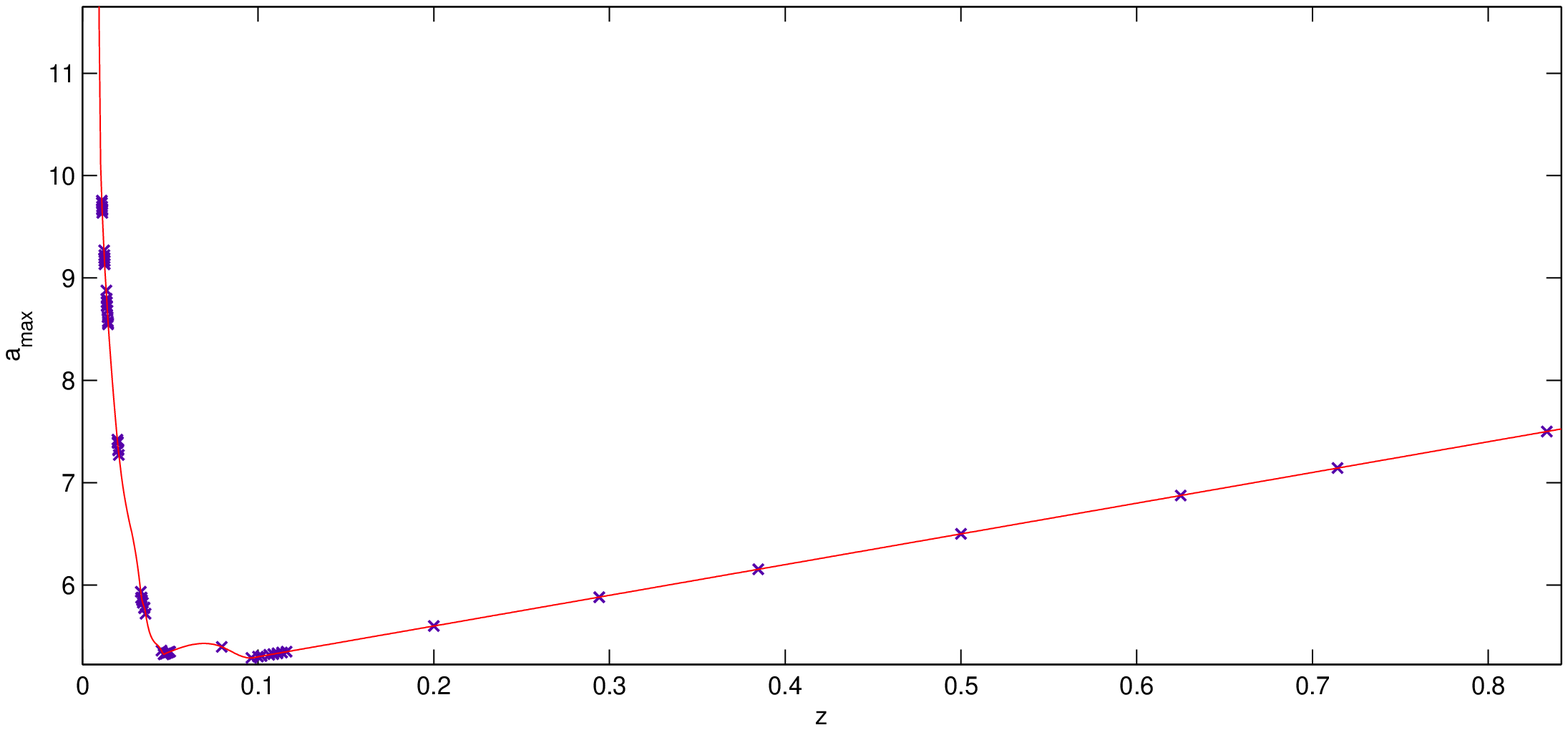, width=12cm}
    \caption{}
  \end{center}
\end{figure}

\begin{figure}[ht]
  \begin{center}
   \epsfig{file=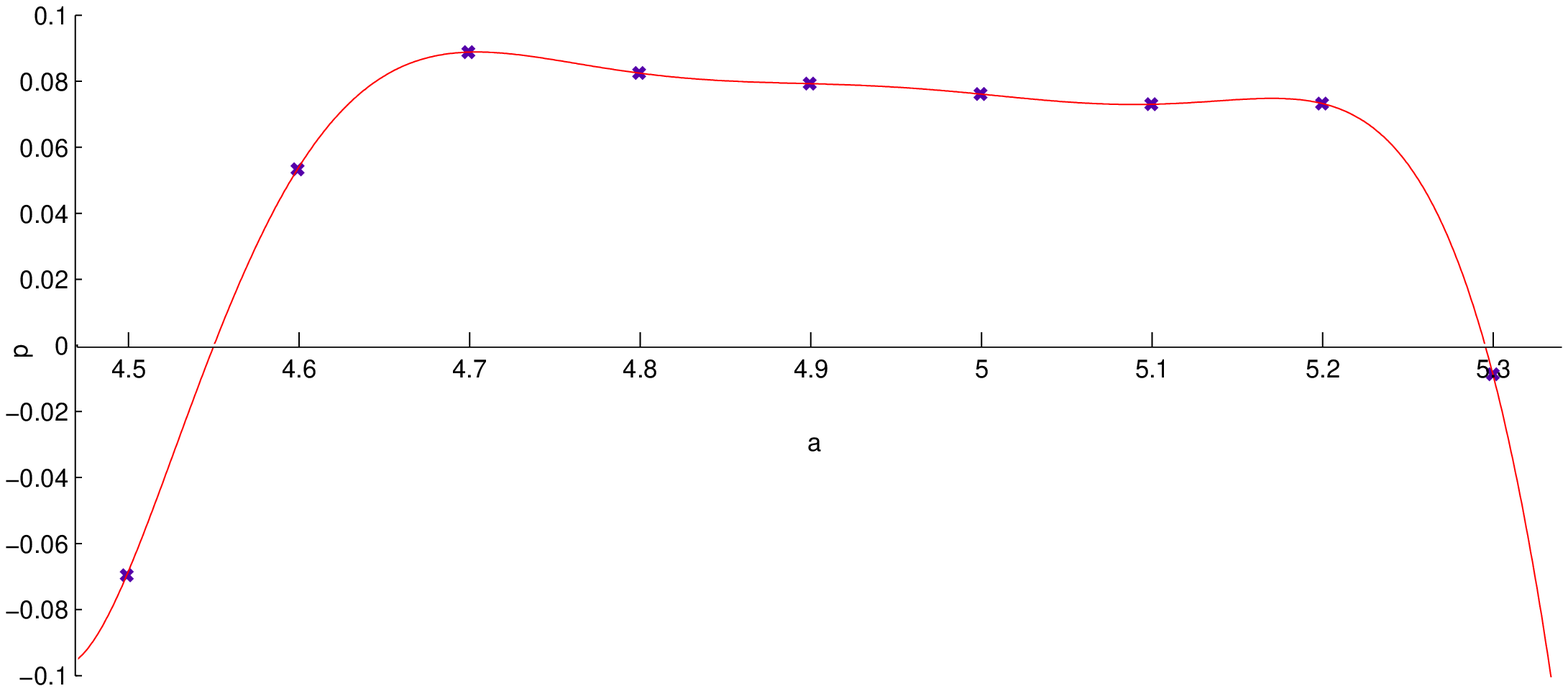, width=12cm}
    \caption{}
  \end{center}
\end{figure}

\begin{figure}
  \begin{center}
   \epsfig{file=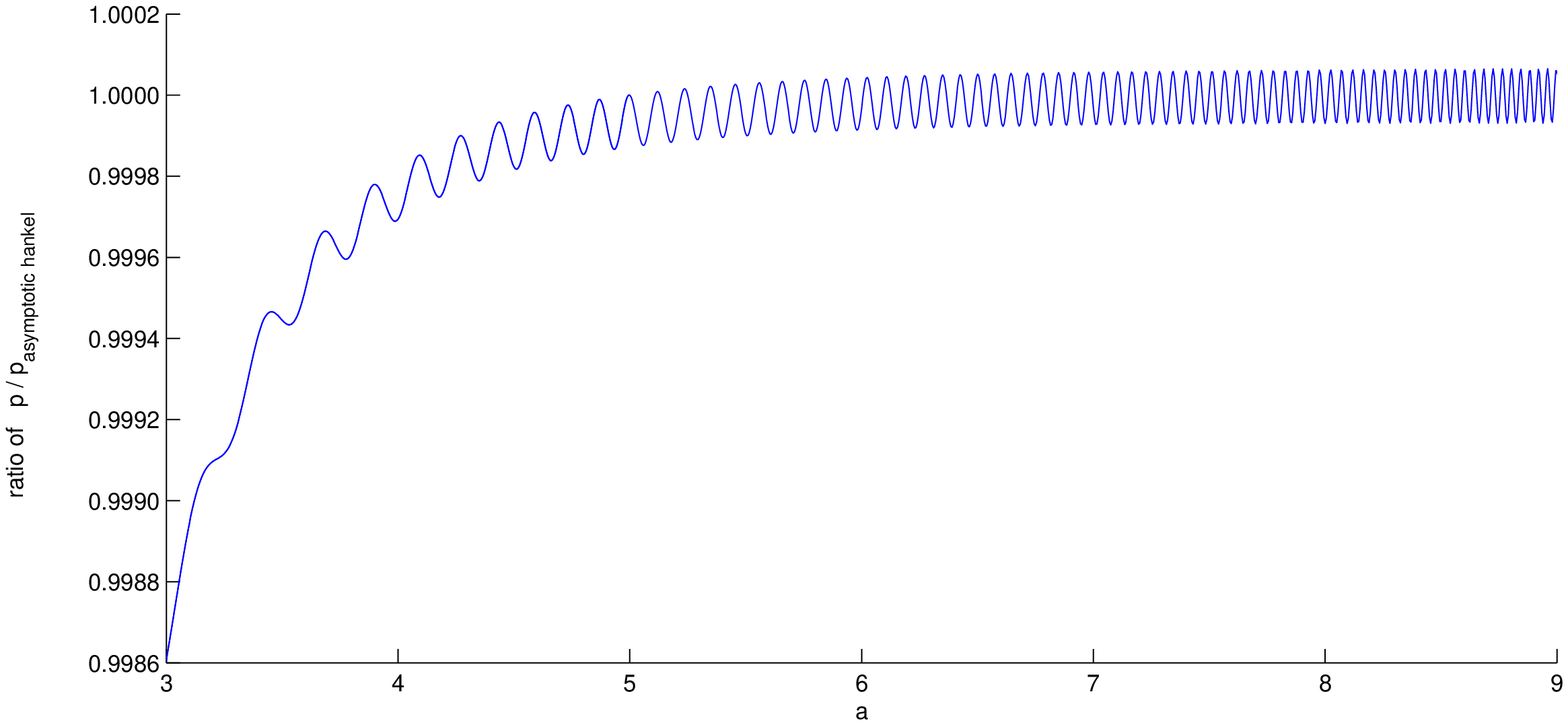, width=12cm}
    \caption{}
  \end{center}
\end{figure}

\begin{figure}
  \begin{center}
   \epsfig{file=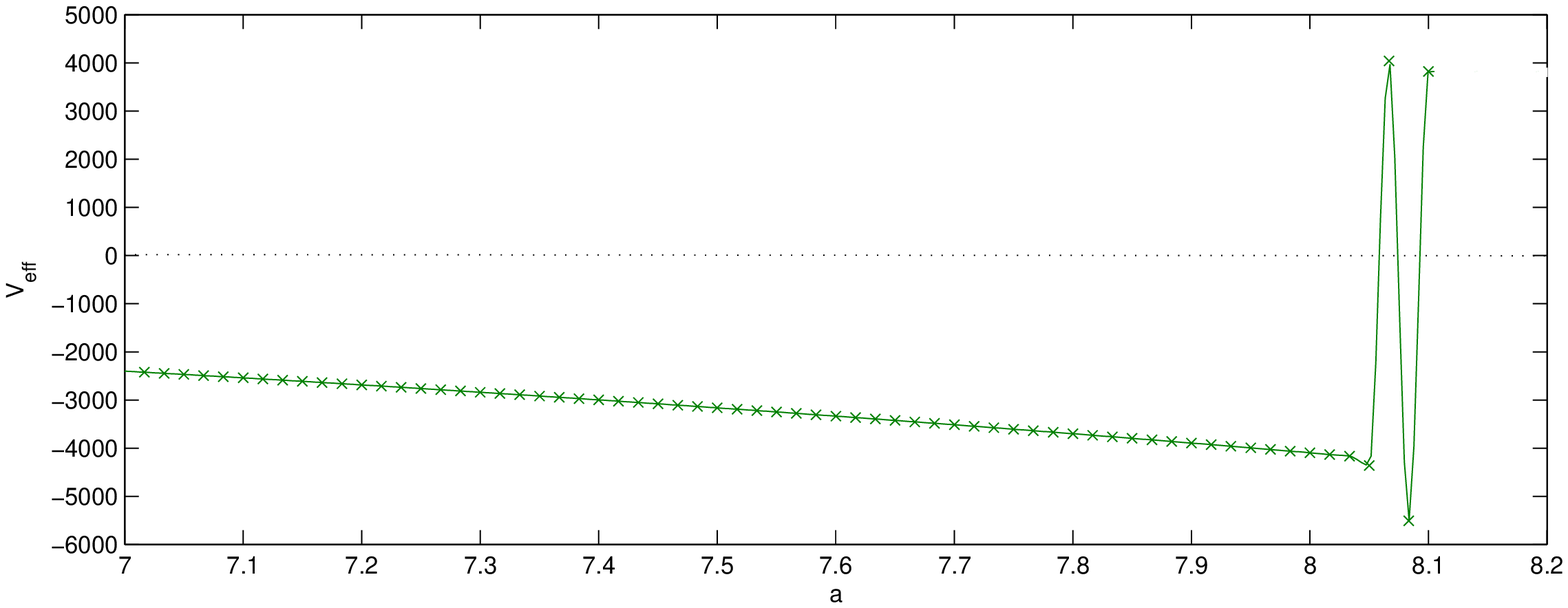, width=12cm}
    \caption{}
  \end{center}
\end{figure}

\begin{figure}
  \begin{center}
   \epsfig{file=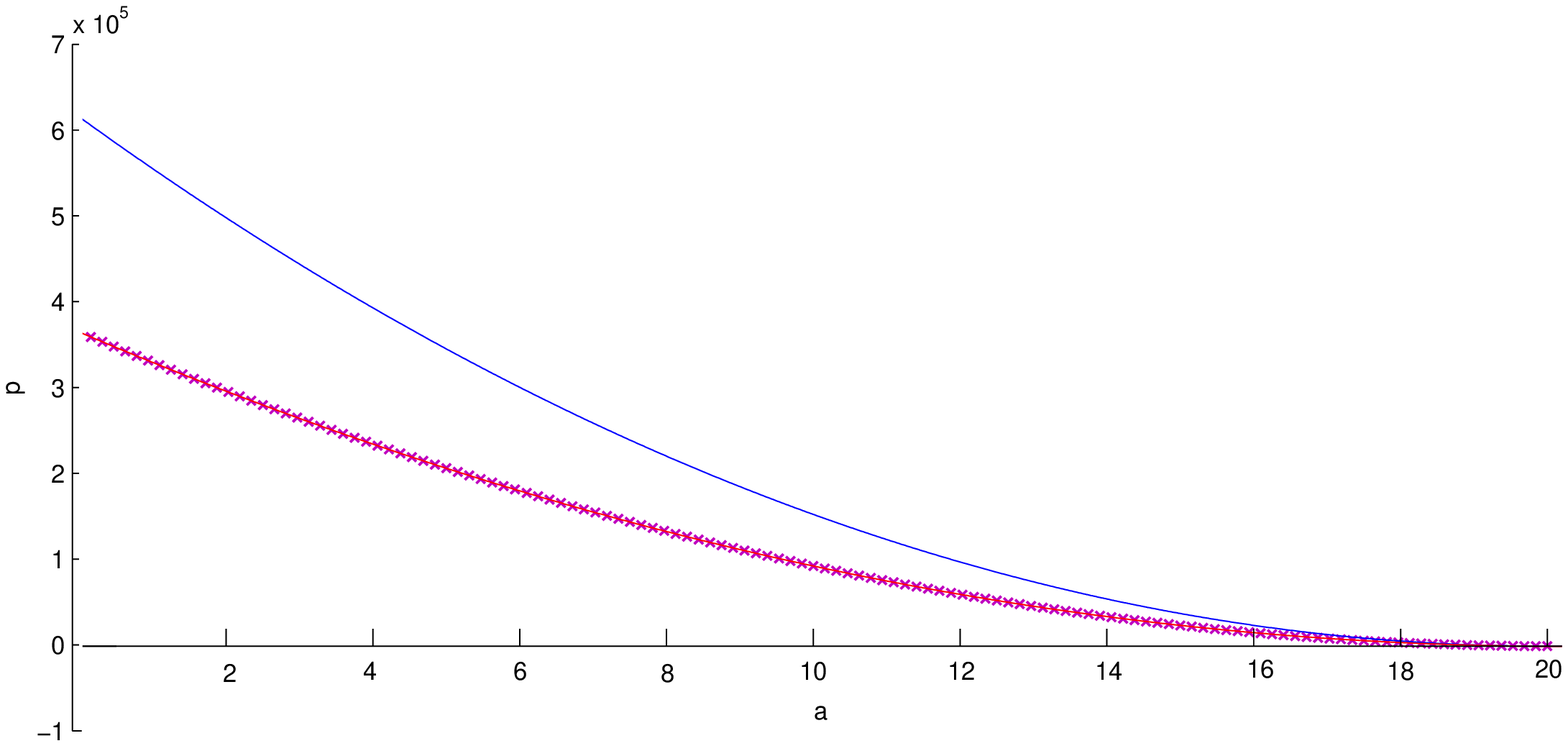, width=12cm}
    \caption{}
  \end{center}
\end{figure}

\begin{figure}
  \begin{center}
   \epsfig{file=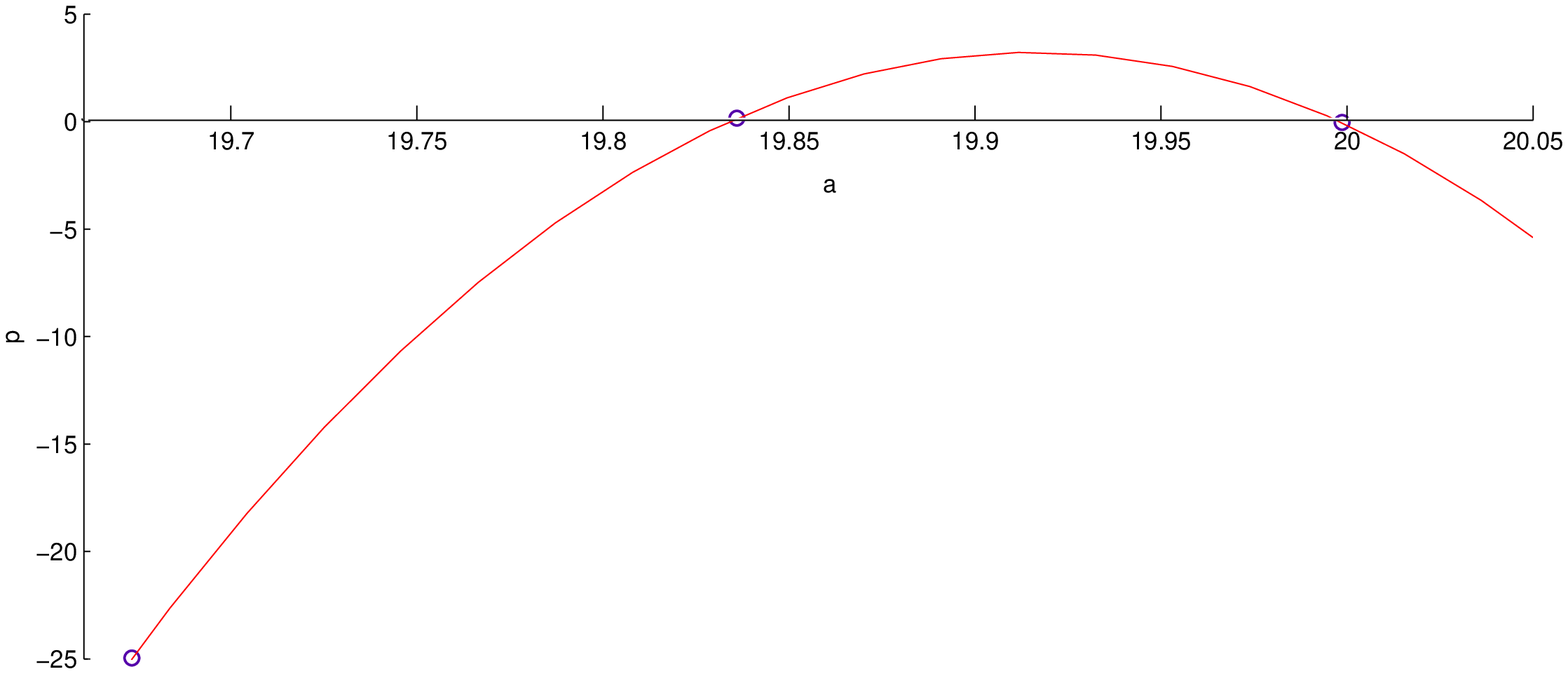, width=12cm}
    \caption{}
  \end{center}
\end{figure}

\end{document}